\documentclass[conference]{IEEEtran}

\newcommand{\ie}{i.e., }
\newcommand{\eg}{e.g., }
\newcommand{\cf}{cf. }
\makeatletter
\newcommand{\etal}{~et~al\@ifnextchar.{}{.\@}}
\newcommand{\etc}{etc\@ifnextchar.{}{.\@}}
\makeatother

\newcommand{\afblock}[1]{\noindent{\textbf{#1}}}

\usepackage{prettyref}
\newrefformat{chap}{Chapter~\ref{#1}}
\newrefformat{sec}{Sec.\,\ref{#1}}
\newrefformat{sub}{Sec.\,\ref{#1}}
\newrefformat{subsec}{Sec.\,\ref{#1}}
\newrefformat{ssub}{Sec.\,\ref{#1}}
\newrefformat{eq}{Eq.~(\ref{#1})}
\newrefformat{fig}{Fig.~\ref{#1}}
\newrefformat{plot}{Fig.~\ref{#1}}
\newrefformat{ex}{Example~\ref{#1}}
\newrefformat{list}{List~\ref{#1}}
\newrefformat{tab}{Table~\ref{#1}}
\newrefformat{enum}{Case~(\ref{#1}.)}
\newrefformat{def}{Definition~\ref{#1}}
\newcommand{\pref}[1]{\prettyref{#1}}

\usepackage[inline]{enumitem}
\SetEnumitemKey{:roman}{label=(\roman*)}
\SetEnumitemKey{:alpha}{label=(\alph*)}
\SetEnumitemKey{:arabic}{label=(\arabic*)}
\SetEnumitemKey{:arabicbold}{label=\textbf{(\arabic*)}}
\SetEnumitemKey{:embeddedtriangle}{label={$\blacktriangleright$},noitemsep,topsep=-1\parskip}

\usepackage{circledsteps}

\usepackage{lipsum}

\usepackage{tikz}

\newcommand*\rectangled[1]{\tikz[baseline=(char.base)]{
    \node[shape=rectangle,draw,inner sep=2pt] (char) {#1};}}

\usepackage{acro}
\makeatletter
\@ifpackagelater{acro}{2017/02/17}{\@ifpackagelater{acro}{2019/09/23}{}{
    \let\IDeclareAcronym\DeclareAcronym
    \renewcommand{\DeclareAcronym}[2]{%
        \IDeclareAcronym{#1}{%
        #2,foreign-plural={}
        }
    }
}}{}

\usepackage{siunitx}
\usepackage{xfp}
\DeclareSIUnit{\noop}{\kern 0pt}
\newcount\tmpnum
\def\store#1#2{\tmpnum=0 \edef\tmp{#1}\storedataA#2\end}
\def\storedataA#1{\advance\tmpnum by1
    \ifx\end#1\else
    \expandafter\def\csname data\tmp\the\tmpnum\endcsname{#1}%
    \expandafter\storedataA\fi
}

\def\assert#1{\ifthenelse{#1}{}{\errmessage{ASSERT FAIL}}}
\def\getarrdata[#1]#2{\ifcsname data#2#1\endcsname\csname data#2#1\endcsname\else\errmessage{UNSET #2#1}\fi}

\def\getarr[#1]#2{\getarrdata[#1]{#2}}
\def\get#1{\getarr[1]{#1}}

\def\roundprefixprefix[#1]#2{\SI[scientific-notation = engineering, exponent-to-prefix = true, round-mode=places,round-precision=#1]{#2}{\noop}}

\def\roundprefix[#1]#2{\ifthenelse {#2 < 1000}{#2\,\,~}{\roundprefixprefix[#1]{#2}}}

\def\round[#1]#2{\SI[round-mode=places,round-precision=#1,round-integer-to-decimal]{#2}{\noop}}

\def\rgetarr[#1]#2{\ifthenelse{\equal{\getarr[#1]{#2}}{ }}{}{\roundprefix[2]{\getarr[#1]{#2}}}}
\def\ngetarr[#1]#2{\ifthenelse{\equal{\getarr[#1]{#2}}{ }}{}{\SI[group-separator={\,}, group-minimum-digits=4]{\getarr[#1]{#2}}{\noop}}}

\def\calcpercprec#1#2#3{$\sim$\SI[round-mode=places,round-precision=#3]{\fpeval{((#1)*1.0)/((#2)*1.0)*100}}{\percent}}

\def\calcpercprec[#1]#2#3{$\sim$\SI[round-mode=places,round-precision=#1]{\fpeval{((#2)*1.0)/((#3)*1.0)*100}}{\percent}}
\def\calcpercprecnosim[#1]#2#3{\SI[round-mode=places,round-precision=#1]{\fpeval{((#2)*1.0)/((#3)*1.0)*100}}{\percent}}

\DeclareAcronym{INC}{
  short         = INC,
  long          = in-network computing
}

\DeclareAcronym{MPC}{
  short         = MPC,
  long          = model predictive control
}

\DeclareAcronym{PND}{
  short         = PND,
  long          = programmable network device
}

\DeclareAcronym{PLC}{
  short         = PLC,
  long          = programmable logic controller
}

\DeclareAcronym{IO}{
  short         = I/O,
  long          = input-output,
  first-style   = short
}

\DeclareAcronym{IoP}{
  short         = IoP,
  long          = Internet of Production
}

\DeclareAcronym{PI}{
  short         = PI,
  long          = proportional integral
}

\store{03_io_f1}{{1.0}}
\store{03_io_recall}{{0.99}}
\store{03_io_precision}{{1.0}}
\store{03_io_accuracy}{{1.0}}
\store{03_warn_f1}{{0.98}}
\store{03_warn_recall}{{0.95}}
\store{03_warn_precision}{{1.0}}
\store{03_warn_accuracy}{{0.98}}
\store{03_error_f1}{{0.98}}
\store{03_error_recall}{{1.0}}
\store{03_error_precision}{{0.97}}
\store{03_error_accuracy}{{0.98}}

\store{08_io_f1}{{1.0}}
\store{08_io_recall}{{1.0}}
\store{08_io_precision}{{1.0}}
\store{08_io_accuracy}{{1.0}}
\store{08_warn_f1}{{0.99}}
\store{08_warn_recall}{{1.0}}
\store{08_warn_precision}{{0.98}}
\store{08_warn_accuracy}{{1.0}}
\store{08_error_f1}{{1.0}}
\store{08_error_recall}{{1.0}}
\store{08_error_precision}{{1.0}}
\store{08_error_accuracy}{{1.0}}

\newcommand{\tone}{T$_{1}$}
\newcommand{\ttwo}{T$_{2}$}
\newcommand{\tthree}{T$_{3}$}
\newcommand{\tfour}{T$_{4}$}

\newcommand{\vone}{V$_{1}$}
\newcommand{\vtwo}{V$_{2}$}
\newcommand{\vmone}{V$_{M1}$}
\newcommand{\vmtwo}{V$_{M2}$}

\newcommand{\pone}{P$_{1}$}

\newcommand{\lone}{L$_{1}$}

\newcommand{\lthree}{L$_{3}$}
\newcommand{\lfour}{L$_{4}$}

\IEEEoverridecommandlockouts
\usepackage{cite}
\usepackage{amsmath,amssymb,amsfonts}
\usepackage{algorithmic}
\usepackage{graphicx}
\usepackage{textcomp}
\usepackage{xcolor}
\usepackage[pdfsubject={https://folks.i4.de/docs/ojflvjkdslfsl}]{hyperref}
\def\BibTeX{{\rm B\kern-.05em{\sc i\kern-.025em b}\kern-.08em
    T\kern-.1667em\lower.7ex\hbox{E}\kern-.125emX}}
\usepackage{relsize}
\usepackage{booktabs}

\newcommand{\oursystem}{\emph{CIVIC}}
\newcommand{\oursystemlongbf}{\textbf{C}ontinuous \textbf{I}n-situ \textbf{V}alidation of \textbf{I}ndustrial \textbf{C}ontrol models}
\newcommand{\oursystemlongemph}{\emph{C}ontinuous \emph{I}n-situ \emph{V}alidation of \emph{I}ndustrial \emph{C}ontrol models}

\newcommand\copyrighttext{%
  \footnotesize \copyright~IEEE, 2024. Personal use of this material is permitted. 
  Permission from IEEE must be obtained for all other uses, in any current or future media, including reprinting/republishing this material for advertising or promotional purposes, creating new collective works, for resale or redistribution to servers or lists, or reuse of any copyrighted component of this work in other works. DOI: \url{https://doi.org/10.1109/ICPS59941.2024.10639999}}
  \newcommand\copyrightnotice{%
  \begin{tikzpicture}[remember picture,overlay]
  \node[anchor=south,yshift=10pt] at (current page.south) {\fbox{\parbox{\dimexpr\textwidth-\fboxsep-\fboxrule\relax}{\copyrighttext}}};
  \end{tikzpicture}%
  }

\begin{document}

\title{In-Situ Model Validation for Continuous Processes Using In-Network Computing}

\author{
  \IEEEauthorblockN{
    Ike Kunze\IEEEauthorrefmark{1},
    Dominik Scheurenberg\IEEEauthorrefmark{2},
    Liam Tirpitz\IEEEauthorrefmark{3},
    Sandra Geisler\IEEEauthorrefmark{3},
    Klaus Wehrle\IEEEauthorrefmark{1}
  }

  \IEEEauthorblockA{
    \IEEEauthorrefmark{1}\textit{Communication and Distributed Systems} $\cdot$
    \{kunze, wehrle\}@comsys.rwth-aachen.de
  }
  \IEEEauthorblockA{
    \IEEEauthorrefmark{2}\textit{Institute of Automatic Control} $\cdot$
    d.scheurenberg@irt.rwth-aachen.de\\
  }
  \IEEEauthorblockA{
    \IEEEauthorrefmark{3}\textit{Data Stream Management and Analysis} $\cdot$
    \{tirpitz, geisler\}@dbis.rwth-aachen.de\\
    All authors are affiliated with \textit{RWTH Aachen University}, Aachen, Germany
  }
}

\maketitle

\begin{abstract}
The advancing industrial digitalization enables evolved process control schemes that rely on accurate models learned through data-driven approaches.
While they provide high control performance and are robust to smaller deviations, a larger change in process behavior can pose significant challenges, in the worst case even leading to a damaged process plant.
Hence, it is important to frequently assess the fit between the model and the actual process behavior.
As the number of controlled processes and associated data volumes increase, the need for lightweight and fast reacting assessment solutions also increases.
In this paper, we propose \oursystem{}, an in-network computing-based solution for \oursystemlongemph{}.
In short, \oursystem{} monitors relevant process variables and detects different process states through comparison with a priori knowledge about the desired process behavior.
This detection can then be leveraged to, \eg shut down the process or trigger a reconfiguration.
We prototype \oursystem{} on an Intel Tofino-based switch and apply it to a lab-scale water treatment plant.
Our results show that we can achieve a high detection accuracy, proving that such monitoring systems are feasible and sensible.
\end{abstract}

\section{Introduction}
\label{sec:introduction}
\copyrightnotice
Ongoing initiatives, such as Industry 4.0~\cite{Lasi:BISE:Industry40} or the \acl{IoP}~\cite{Pennekamp:ICPS:InfrastructureIoP}, further fuel the ongoing digitalization of production industries and, \eg enable the creation of new knowledge.
One beneficiary of such data-driven approaches is industrial process control~\cite{Rueppel:IoPBuch23:B2II}.
For example, \ac{MPC} can effectively control continuous production processes, 
but requires precise system models to first predict future behavior and then apply suitable actions~\cite{Morari:CCE99:MPC}.

In practice, small deviations between the model and the real process usually only (slightly) reduce the control performance of \ac{MPC}.
In contrast, larger differences can cause significant misbehavior of the control and ultimately lead to considerable damage to the process plant.
To ensure the use of a sufficiently accurate process model, it is important to detect and handle any potentially dangerous deviations as early as possible~\cite{Badwe:JPC09:ImpactModel}.

Modern decentralized  manufacturing, however, often outsources process control to powerful cloud environments as conventional field-based devices are often insufficient for advanced control and as deploying dedicated local compute resources is cost-intensive~\cite{Cao:TII2021:ICPSCloudEdge}.
The additional latency to the cloud can become problematic for \ac{MPC} as it delays potential responses to differences between the model and the process~\cite{Rueth:NetCompute2018:InNetworkControl}.
Hence, there is a strong need for a dedicated, but lightweight monitoring system on the edge that can detect such changes, \eg by constantly monitoring the current process state and comparing it with the expected behavior to quickly trigger adequate responses.

Based on related work, we identify \ac{INC} as a promising solution platform.
In short, \ac{INC} uses the capabilities of modern networking devices to move computations to the data path, allowing earlier processing at higher processing rates compared to end-host systems.
Previous work has already demonstrated the general applicability of \ac{INC} to industrial scenarios~\cite{Kunze:ICPS2021:ApplicabilityINC} and to the control of robot arms~\cite{Cesen:NetSoft20:RobotControl,Laki:NSDI22:VelocityControl,Wang:MobiSys23:KneeJerk}.
While the former approach only performs single operations per packet, the latter works rely on precise models of the underlying processes, similar to \ac{MPC}.
Hence, they are likely also affected by changes in process behavior, but do not yet explicitly detect or handle these cases.

To address this gap, we propose \oursystem{}, a \oursystemlongbf{}.
\oursystem{} monitors real process communication, extracts relevant sensor and control information, and aggregates this information to describe the current process state.
It then checks this state against an expected system model with the goal of identifying relevant deviations.
We prototype \oursystem{} on an Intel Tofino-based switch and demonstrate its practicability by applying it to a lab-scale water treatment plant.
Our experiments show that \oursystem{} can reliably detect deviations from the expected process behavior.
Overall, we contribute the following:
\begin{itemize}
    \item We identify the need for validating the system models of continuous industrial processes at run-time.
    \item We design \oursystem{} which monitors the communication of industrial processes to assess their system behavior.
    \item We demonstrate the efficacy of \oursystem{} at the example of a lab-scale water treatment plant and show that it can efficiently assess the fit of the process model.
\end{itemize}

\afblock{Structure.}
In \pref{sec:background}, we introduce continuous industrial processes and discuss related work on \ac{INC}.
\pref{sec:design} then presents the design of \oursystem{} which we implement on an Intel Tofino-based switch and apply to a water treatment plant in \pref{sec:impl_model_factory}.
We evaluate \oursystem{} in \pref{sec:evaluation}.
\pref{sec:conclusion} concludes the paper.

\acresetall

\section{Controlling Continuous Industrial Processes}
\label{sec:background}
Modern industrial ecosystems are heterogeneous and decentralized environments that enable dynamic task allocation to diverse resources.
Leveraging this flexibility, the control of many continuous industrial processes, as, \eg used in the chemical industry, is increasingly spatially decoupled and moved away from the plants to optimally utilize powerful hardware and satisfy ever growing quality demands.
Yet, this spatial separation also raises challenges, \eg by inducing latencies that delay control responses.
Hence, it is still sensible to push \emph{some} lightweight, time-sensitive operations to the network edge.
In the following, we present continuous processes and associated challenges in more detail and further discuss how research addresses this field.

\begin{figure}[t]
\centering
\includegraphics[width=0.48\textwidth]{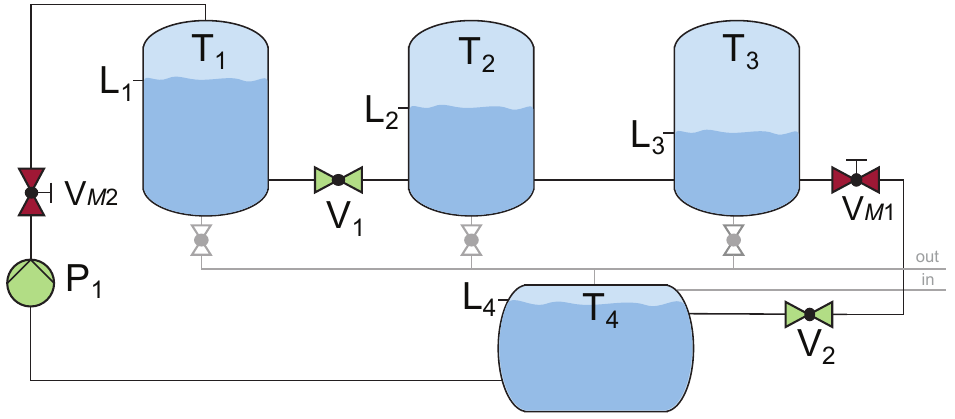}
\caption{The coupled tank system uses pump \pone{}, and valves \vone{}/\vtwo{} to control the water levels \lone{}-\lthree{}.
The manual valves \vmone{}/\vmtwo{} can emulate faults. 
}
\label{fig:tank_schematic}
\end{figure}

\afblock{Continuous industrial processes.}
Continuous processes transform raw materials into finished products without interruption.
For example, chemical plants utilize such processes for reactions or refinements while water treatment plants continuously filter or disinfect water.
\pref{fig:tank_schematic} shows a simple coupled tank system that provides water to a larger continuous water treatment pipeline~\cite{Scheurenberg:AIM22:CoupledTank} for which the water levels \lone{}-\lthree{} in tanks \tone{}-\tthree{} need to be controlled.
In all examples, continuous and precise process control is needed to ensure operational efficiency and consistent product quality.

\afblock{Controlling and monitoring continuous processes.}
In practice, feedback control is the most prominent control scheme: 
the process "output", \eg the water levels \lone{}-\lthree{}, is continuously monitored while "input" parameters, such as pump speed \pone{} or the valves \vone{} and \vtwo{}, are adjusted based on the difference between the desired and actual output.
\Ac{PI} control is a simple feedback control that is easy to implement and sufficient for many processes.
However, it is bad at handling nonlinearities or constraints, and lacks predictive capabilities, \ie it only responds to the current state without considering future behavior, such that additional mechanisms may be needed to prevent system limit violations.

\afblock{\Acf{MPC}.}
\Ac{MPC} addresses several limitations of \ac{PI} control and is preferred for complex and dynamic processes that require an adaptable and more precise control.
In short, \ac{MPC} solves an optimization problem at iterative time steps to calculate the optimal plant input given certain constraints.
For this, it predicts the process behavior over a given horizon based on a process model and then considers the current state, the predicted future states, and desired objectives to determine optimal control actions with respect to a cost function~\cite{Morari:CCE99:MPC}. 
However, its effectiveness depends on the accuracy of the model: if it poorly represents the actual process dynamics, the control performance can suffer.
Additionally, wear of or damage to production plants often change the actual behavior.
Hence, regular model validation and refinement are important for maintaining the accuracy of the process model.

\afblock{Validating the process model in situ.}
Feedback control in general already includes monitoring the controlled variables.
Validating the process model further requires monitoring other process parameters that provide additional information on the process state and whether changes occurred~\cite{Yen:IEM19:CPSMonitoring}.
In particular, smaller changes might only require model updates while larger changes might lead to unsafe states or behavior as, \eg a severe malfunction of pump \pone{} might require a system shutdown.
Hence, such safety-critical states need to be detected as quickly as possible.
However, existing approaches for plant-wide monitoring of large-scale industrial production plants require powerful hardware and perform time-intensive computations.
Since these solutions rely on sensor fusion and cause large data volumes that need to be transmitted and processed~\cite{Glebke:hicss2019:integratedData}, they delay any potential reaction, especially when spatially separated from the process~\cite{Beyca:TASE16:SensorFusion, Gao:TIE15:FaultDiagnosis}.
As a new solution space, research investigates monitoring and analyzing certain process characteristics close to the process on \acp{PND}. 

\afblock{\Ac{INC} and related work.}
\Ac{INC} refers to deploying functionality on \acp{PND} that can process large data volumes at high rates but are often limited in their processing capabilities as, \eg multiplications are typically only possible with a fixed second factor and involving powers of two~\cite{Kunze:IM2021:AQMTofino}.
Yet, their privileged position on the data path and close to the data source makes \acp{PND} an attractive target for processing massive industrial data streams with strict latency requirements.
Approaches on protocol level, \eg utilize the structure of industrial protocols to filter out messages and reduce traffic~\cite{Gyorgyi:NetSoft2021:InNetworkTrafficReduction}, but ignore the actual payload. 
Payload-focused proposals inspect individual packets to perform simple computations, such as coordinate transformations~\cite{Kunze:ICPS2021:ApplicabilityINC} or data stream processing tasks~\cite{Sankaran:NetSoft21:INCScientificData}.
PNDs can also be used to react to data in individual packets, \eg by issuing alerts (if monitored sensor readings violate thresholds~\cite{Atutxa:Sensor2021:IndLowLatency}) or even stopping machines~\cite{Cesen:NetSoft20:RobotControl}.
While computations and reactions on the level of individual sensor readings already demonstrate the usefulness of PNDs for industrial data processing in general and the detection of error cases specifically, industrial processes are often characterized by long-term dependencies, across series of readings.
Laki et al.~\cite{Laki:NSDI22:VelocityControl} use a PND to plan and continuously control the trajectory of a robot by implementing PID-like control functionality.
Kunze et al.~\cite{Kunze:ISIE2021:SignalDetection} monitor cyclic manufacturing processes to detect instabilities by taking advantage of the repeating signal pattern of this class of processes.
However, this approach is not feasible for continuous processes, which we consider in this paper.

In summary, there is a strong need for dynamic approaches capable of validating models of continuous industrial processes at run-time.
While \ac{INC} is a potential fit, no work has yet addressed this topic.
Hence, in this paper, we propose \oursystem{}, a \oursystemlongbf{} which compares process behavior to process models using \ac{INC} to enable model refinement and emergency reactions.

\section{\oursystemlongbf{} (\oursystem{})}
\label{sec:design}

A continuous in-situ validation of process models can help to ensure a high process control quality.
In particular, control schemes heavily relying on these models, such as \ac{MPC}, can significantly benefit while validation results can also be used for triggering warnings or emergency shutdowns if there is a significant system misbehavior.
\oursystem{} uses \ac{INC} to provide such an in-situ validation for continuous production processes.

\begin{figure}[t]
\centering
\includegraphics[]{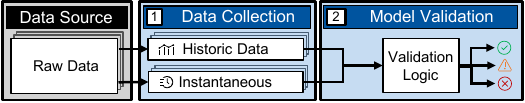}
\caption{\oursystem{} consists of two components: its \protect\rectangled{1} data collection unit collects instantaneous and long-term process information which the \protect\rectangled{2} model validation unit uses to assess the fit of the process to the given model.}
\label{fig:concept}
\end{figure}

\afblock{Design overview.}
In essence, \oursystem{} continuously monitors periodic process communication, extracts relevant information for monitored components, and validates the process model, \eg via known dependencies.
Conceptually, \oursystem{} consists of two components as visualized in \pref{fig:concept}:
\begin{enumerate*}[label=\protect\rectangled{\arabic*}]
  \item a data collection unit, and 
  \item a model validation unit.
\end{enumerate*}
In the following, we present the two components in more detail.

\subsection{Data Collection}
\label{sub:design_data_collection}
The data collection unit monitors periodic process communication to collect information on all process components of interest, potentially for multiple processes at the same time.
Given that production plants grow in size and often feature many processes, each consisting of numerous sensors and components, one particular challenge for any data collection unit is the processing of increasingly large data volumes~\cite{Glebke:hicss2019:integratedData}.
Addressing this challenge, we envision to deploy the data collection unit on \acp{PND} as they are capable of handling traffic at speeds of Tbps and are specifically designed for extracting specific (bit-level) information on a per-packet basis~\cite{intel:tofino}.

\afblock{What information to extract?}
The scope and form of the extracted information depend on the monitored process and the desired validation steps.
In particular, one might be able to validate some aspects of a process using instantaneous data, \eg values from a single sensor reading, while other aspects might require longer term dependencies.
For this purpose, \oursystem{} uses registers to provide both capabilities, \ie \emph{instantaneous} information as well as a \emph{history} of up to $n$ previous readings where $n$ can differ for different information sources or even for the same information source when used in different processes.
Based on the collected information, \oursystem{} then validates the underlying system model.

\subsection{Model Validation}
\label{sub:design_model_validation}
The model validation unit checks the monitored information for known properties to detect relevant deviations.
Accounting for the diversity of continuous processes, we do not propose a single one-fits-all solution, but instead aim to provide different validation modules that are applicable to different scenarios.
Hence, we next non-exhaustively illustrate the possible range of mechanisms with \emph{some} possible forms of validation.

\afblock{Validation logic.}
A simple validation is to check specific system thresholds or combinations thereof, such as minimum or maximum fill levels of water tanks, pump speeds, or valve positions (\cf\pref{fig:tank_schematic}).
More advanced validation schemes could combine the simple checks with longer term assessments, \eg concerning the rate of water increase or decrease in the tanks, to allow for a broader understanding of the system.
For example, if \pone{} is turned off and \vone{} is closed, we would not expect any changes to the fill level \lone{} while opening \vone{} should induce a known water discharge profile.
Overall, concrete validation steps are highly process-dependent. 

\afblock{Validation complexity.}
Validation logic can also significantly differ in complexity.
The mentioned threshold checks, \eg only require simple comparisons while checking for change \emph{rates}, \eg requires computing slopes.
At this point, the computational capabilities available on \acp{PND} influence which forms of validation are possible.
Hence, when aiming to deploy \oursystem{} for a specific process, it is important to consider which forms of validation are desired and which \acp{PND} are available.
However, even if some operations do not map to the fast data plane, more expressive operations can be executed on the slower control plane at the cost of higher delays~\cite{Kunze:ISIE2021:SignalDetection}.

\afblock{Severity levels.}
Based on the validation, \oursystem{} needs to take appropriate action.
Minor deviations from the process model typically only slightly impact the system such that immediate action is beneficial but not critically important.
Hence, raising warnings is often enough, \eg to trigger a reconfiguration.
In contrast, larger deviations can hint at imminent critical damage and the system should be immediately brought to a safe state.
Overall, we identify at least three validity levels that \oursystem{} needs to distinguish:
\begin{enumerate*}[label=(\roman*)]
  \item normal operation,
  \item small deviations (``warning''), and 
  \item large deviations (``error'').
\end{enumerate*}

In the following, we demonstrate the potential of our considerations by applying \oursystem{} to a lab-scale water treatment plant using a real \ac{PND} as our deployment platform.

\section{Monitoring a Water Treatment Plant}
\label{sec:impl_model_factory}

We implement \oursystem{} in P4 on an Intel Tofino switch~\cite{intel:tofino} and use it to monitor a lab-scale water treatment plant used for research~\cite{Scheurenberg:AIM22:CoupledTank} and teaching purposes. 
In the following, we present the plant and discuss how we deploy \oursystem{} using the data collection and model validation modules shown in \pref{fig:tofino_concept}.

\subsection{Water Treatment Plant}
\label{sub:water_treatment_plant}
Our water treatment plant comprises several stations, each with dedicated equipment to implement a specific continuous operation, such as water level control, heating, or pH adjustment.
In this paper, we focus on the first station of the plant: 
a coupled tank system as depicted schematically in \pref{fig:tank_schematic}.
It consists of four distinct tanks (\tone{}-\tfour{}) and the primary control aims to achieve and sustain specific water levels (\lone{}-\lthree{}) within the upper tanks.
For this, one pump (\pone{}) and two valves (\vone{},\vtwo{}) are controlled automatically via an \ac{MPC} while additional valves \vmone{}{} and \vmtwo{}{} can be adjusted manually.

\afblock{System behavior.}
The upper tanks \tone{}-\tthree{} are positioned at the same height.
Since water not only flows into the succeeding tank but also affects the preceding tank, \eg via backflow due to sweeping movements, \tone{}-\tthree{} are interdependent.
We control the water flow into \tone{} via the speed of pump \pone{} and the water flows \tone{} to \ttwo{} and \tthree{} to \tfour{} via pneumatic ball valves \vone{} and \vtwo{}, each equipped with position controllers.
Overall, the behavior of the plant is characterized by the (nonlinear) interaction of water levels, valve states, and pump activity.

\afblock{System model.}
The properties of the coupled tank system have been approximated via empirical investigations for which we have performed a dead-time measurement with regard to \pone{} and the impact of \vone{} and \vtwo{}, yielding nonlinear characteristic maps of the components and resulting in a higher-order nonlinear system.
However, it is possible to create a linear system model by linearization at suitable operating points, which can describe a large part of the operating range sufficiently well.

\subsection{Data Collection}
\label{sub:impl_data_collection}
\oursystem{} extracts the current status of each component of the water treatment plant from periodic process control messages.
In particular, we track the current water levels (\lone{}-\lfour{}), the valve states (\vone{},\vtwo{}), and the pump speed of \pone{} (\cf\pref{fig:tofino_concept}).

\afblock{Data collection on Tofino.}
We extract individual values from control messages via Tofino’s parser, treating the entire message as a packet header.
These instantaneous values are directly available for model validation.
For the main control variables \lone{}-\lfour{}, we keep a longer history via dedicated ring buffers $R_{i}$ for each variable.
They have different sizes and are each implemented using two registers: one holding a pointer to the current index, the other storing the actual data (\cf\cite{Kunze:EPIQ2021:SpinTracker}).

\subsection{Model Validation}
\label{sub:impl_model_validation}
\oursystem{} uses instantaneous and long-term data provided by the data collection unit for validating the system behavior.
As the water treatment plant’s model specifically describes the dependencies between different valve states, water levels, and pump speeds, we focus on these aspects for our validation. 

\begin{figure}[t]
	\centering
	\includegraphics[]{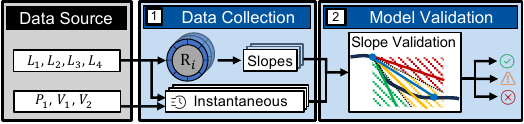}
	\caption{Our prototype \protect\rectangled{1} tracks instantaneous control data (\pone{},\vone{},\vtwo{}) and long-term information on \lone{}-\lfour{}.
  \protect\rectangled{2} We compare the long-term slopes against given references when the instantaneous data matches certain thresholds.
  }
	\label{fig:tofino_concept}
\end{figure}

\afblock{Inhibited water flow.}
Most misbehavior in our plant affects the water flow between different tanks.
In particular, we typically expect specific water movement depending on the overall system state, \ie the fill levels of the tanks, the state of the valves, and the speed of the pump.
Hence, we set up \oursystem{} to monitor the water movement between the tanks over longer timespans while we use instantaneous information on the control parameters to assess the actuator state (\cf\pref{fig:tofino_concept}).

\afblock{Monitoring water flows.}
There are different possible strategies for monitoring the water movement.
We opt for tracking the water levels in each tank over operating windows of $n$ control messages after which we determine volume changes by performing a linear regression on the entire window to obtain a smoothed average volume change.
In our initial empirical system assessment, we have already derived three baseline slopes, one for each of the validity levels defined in \pref{sub:design_model_validation}.
Hence, we can assess the adherence to the model by comparing the slope derived via linear regression with the three validity slopes where we choose the closest one as the validity class.

\afblock{Model validation on Tofino.}
While our assessment logic is straightforward, linear regressions over longer windows are usually infeasible on \acp{PND}.
Hence, we adapt our methodology as follows:
instead of a full regression, we approximate the slope by subtracting the latest value in the window from the oldest one.
We then determine the validity class by comparing the resulting slope with baseline values using a dedicated P4 table with range matches configured with specific ranges between the baseline slopes (shared areas in \pref{fig:tofino_concept}).

\afblock{Decision Logic.}
For this paper, we configure \oursystem{} to only label outgoing packets with the current detection state.
In the future, we envision to feed this information back into the process control, \eg to switch between different available system models depending on the system behavior.

In the following, we evaluate the performance of \oursystem{}.

\section{Evaluation}
\label{sec:evaluation}

\oursystem{} is designed to validate that the behavior of our water treatment plant follows expected behavior.
For evaluating its efficacy, we control the plant assuming nominal operation, deliberately introduce errors in the system, and then evaluate how well and fast \oursystem{} can detect the corresponding deviations in system behavior.
In the following, we present our evaluation methodology in more detail before presenting our results.

\subsection{Methodology}
\label{sub:methodology}
We evaluate our prototype based on real data from our water treatment plant collected during several experiments.

\afblock{Process control.}
In the experiments, we control the plant with a linear \ac{MPC} scheme that we have extensively evaluated in previous work~\cite{Scheurenberg:AIM22:CoupledTank}, allowing for a responsive and precise control of the system.
In short, we linearize the system model (\cf\pref{sub:water_treatment_plant}) at a specific operating point that is partially characterized by the fill levels of the upper tanks (\lone{}-\lthree{}), and then discretize the model for a sampling rate of \SI{10}{\Hz}.

\afblock{Communication setup.}
The coupled tank system is equipped with dedicated sensors and actuators that are connected to an S7 \ac{PLC} serving as a pure data gateway.
The actual control runs on a dedicated computer that is connected via Profinet and communicates via UDP.
We place \oursystem{} between the \ac{PLC} and the control computer such that it can operate on actual real-world control messages.

\afblock{Datasets.}
Our experiments cover different fault states as well as nominal behavior of the water treatment plant.
In particular, we emulate faults of pump \pone{} and of the pipe draining \tthree{} by partially closing the manual valves \vmone{} and \vmtwo{}.
During the experiments, we capture the periodic control traffic from our gateway \ac{PLC} to our control computer which includes all sensor values and the current configuration of the control parameters.
We capture dedicated traces for ``training'' (five runs) and testing (ten runs) in each of the considered scenarios.

\afblock{Evaluation.}
To enable reproducibility, we replay the recorded traffic from a dedicated machine \emph{M$_{1}$} to one Tofino-based switch running \oursystem{}.
The switch performs the validation, labels the traffic accordingly, and then forwards it to machine \emph{M$_{2}$} where it is captured for analysis.
We assess \oursystem{}’s performance by comparing its labeling with the expected results of the scenario, \ie if we do not emulate a fault, \oursystem{} should identify normal system operation.
As \oursystem{} yields unambiguous results, we replay each setting once.

\begin{table}[t]
  \centering
    \caption{Classification scores of \oursystem{} in the clogged pipe setting with adjusted start condition.} 
  \label{tab:table_valve}
  \def\arraystretch{0.5}
  \begin{tabular}{r|c|c|c|c}
    Class         & Precision                   & Recall                  & F1                    & Accuracy \\
    \midrule
    Normal        & \get{03_io_precision}       & \get{03_io_recall}      & \get{03_io_f1}        & \get{03_io_accuracy} \\
    Warning       & \get{03_warn_precision}     & \get{03_warn_recall}    & \get{03_warn_f1}      & \get{03_warn_accuracy} \\
    Error         & \get{03_error_precision}    & \get{03_error_recall}   & \get{03_error_f1}     & \get{03_error_accuracy} \\
  \end{tabular}
  \vspace{-0.25em}
\end{table}

\subsection{Identifying Clogged Pipes}
\label{sub:valve_classification}
A frequent fault in our plant is a clogging of pipes due to deposits which inhibits the possible water flow and worsens process control.
Our first evaluation case covers this scenario.

\afblock{Fault emulation.}
We emulate a clogged pipe by a partial closure of manual valve \vmone{}, inhibiting the water flow from \tthree{} to \tfour{}.
Based on the severity levels defined in \pref{sub:design_model_validation}, we distinguish three cases:
\begin{enumerate*}[label=(\roman*)]
  \item \emph{normal} operation where \vmone{} is completely open,
  \item a \emph{warning} scenario where we slightly close \vmone{} to \SI{30}{\percent}, and 
  \item an \emph{error} scenario where we close \vmone{} to \SI{55}{\percent}, significantly changing the system state.
\end{enumerate*}

\afblock{Validation logic.}
For detecting the clogged pipe, we trigger a water discharge from \tthree{} via \vtwo{} and analyze the corresponding volume changes over an operating window of $20$ sensor intervals.
As a discharge requires a sufficient initial volume in \tthree{} and a sufficiently open valve, we only apply this logic if the volume in \tthree{} (\lthree{}) is larger than \SI{3}{\liter} and if \vtwo{} is at least \SI{80}{\percent} open.
We realize this logic by combining the instantaneous threshold and long-term slope validation primitives of \oursystem{}.

\afblock{Baseline results.}
In a first step, we train \oursystem{}’s slope validation primitive on a training set of five runs collected for an initial volume in \tthree{} of \SI{7.5}{\liter}.
For this, we manually optimize \oursystem{}’s range matches such that we achieve perfect results on the training set.
We then evaluate the corresponding performance on a test set of ten runs.
We find that \oursystem{} still achieves a perfect assessment in these runs, \ie all cases are classified correctly without any false or missed classifications as indicated by classification scores of $1$.
However, the system can have different initial configurations and we have only picked a single one.
Hence, to study \oursystem{}’s broader applicability, we decide to also test a different initial state.

\afblock{Changing the start condition.}
We repeat the experiment but change the initial volume in \tthree{} to \SI{6.5}{\liter} which reduces the slope of the discharge profile due to smaller water pressure.
In this setting, we only collect a test set of ten runs.
\pref{tab:table_valve} shows the classification scores when using \oursystem{} with its configuration from above.
As can be seen, we still achieve a high performance and correctly identify all error cases (indicated by a recall of $1$).
While we misclassify some normal and warning cases as the next higher severity, we argue that being a bit overcautious is desirable from a process safety standpoint. 
Hence, overall, \oursystem{} performs well and its configuration can also cope with smaller system setup changes.

\subsection{Detecting a Failing Pump}
\label{sub:pump_classification}
Another common fault is a failing pump, \eg moving less volume than normal, which we use as our second case.

\afblock{Fault emulation.}
Similar to the clogged pipe, we emulate a fault of \pone{} by partially closing \vmtwo{}, inhibiting the water flow from \tfour{} to \tone{}.
We again use three severity levels:
\begin{enumerate*}[label=(\roman*)]
  \item \emph{normal} operation with \vmtwo{} open,
  \item \emph{warning} where \vmtwo{} is closed to \SI{20}{\percent}, and 
  \item \emph{error} where \vmtwo{} is closed to \SI{40}{\percent}.
\end{enumerate*}  

\afblock{Validation logic.}
Depending on \pone{}’s speed, we expect a corresponding volume increase in \tone{}.
Hence, we validate \pone{} by tracking the slope of \lone{} when filling an almost empty \tone{}.
As this change is slower than the water discharge, we use a longer operating window of $90$ sensor intervals and only perform the validation if \pone{} has a high speed (\pone{} $>$ \SI{99}{\percent}) and if the water level in \tone{} is in a certain corridor (\SI{5}{\liter}$<$\lone{}$<$\SI{8}{\liter}).

\afblock{Results.}
Similar to \pref{sub:valve_classification}, we first train \oursystem{}’s slope validation primitive on a training set of five runs and then evaluate its performance on a test set of ten runs.
\pref{tab:table_pump} shows the classification performance results.
As can be seen, \oursystem{} achieves high classification quality for detecting a failing pump, too.
However, similar to the clogged pipe setting, we again observe a few warning cases that are misclassified as errors, \ie \oursystem{} once more shows overcautious behavior.

\begin{table}[t]
  \centering
    \caption{Classification scores of \oursystem{} in the failing pump setting.} 
  \label{tab:table_pump}
  \def\arraystretch{0.5}
  \begin{tabular}{r|c|c|c|c}
    Class         & Precision                   & Recall                  & F1                    & Accuracy \\
    \midrule
    Normal        & \get{08_io_precision}       & \get{08_io_recall}      & \get{08_io_f1}        & \get{08_io_accuracy} \\
    Warning       & \get{08_warn_precision}     & \get{08_warn_recall}    & \get{08_warn_f1}      & \get{08_warn_accuracy} \\
    Error         & \get{08_error_precision}    & \get{08_error_recall}   & \get{08_error_f1}     & \get{08_error_accuracy} \\
  \end{tabular}
  \vspace{-0.25em}
\end{table}

\subsection{Discussion}
\label{sub:discussion}

\oursystem{} is designed to validate the behavior of continuous processes and our evaluation demonstrates its general capabilities and that it can effectively track a real process.
In the following, we shortly discuss possible extensions of \oursystem{} and how it can be directly embedded into the process control.

\afblock{Embedding \oursystem{} into process control.}
\oursystem{} can detect if processes no longer comply with their expected behavior.
While we currently only label packets, there are different options for integrating \oursystem{} into the process control.
For example, we could analyze different system states upfront and create unique models for each of them.
Based on \oursystem{}’s assessment, we could then quickly change between the models, always choosing the best match.
Similarly, there could be different control parameterizations, \eg conservative and more aggressive ones, and \oursystem{}’s assessment could be used to choose among them.
Finally, more complex embedding could directly leverage the output of \oursystem{} for the process control.

\afblock{Severity levels.}
In this work, we experiment with three basic fault severity levels.
In practice, however, there could be a larger range of such levels, each having a different implication on the overall process health.
Additionally, more levels would also enable more configuration options when embedding \oursystem{} into the process control.
Hence, in future work, we aim to dive deeper into possible fault states.

\afblock{Application scenario.}
In this paper, we apply \oursystem{} to a lab-scale water treatment plant which we can monitor using the slope validation module.
This allows us to verify that our concepts work as intended and can operate on real process communication.
A wider selection of validation modules for \oursystem{} would extend support to a broader range of processes.

\section{Conclusion}
\label{sec:conclusion}

Modern industrial control approaches increasingly rely on complex models of the controlled systems.
While such data-driven concepts enable an efficient and highly performant control as long as the model accurately reflects the system behavior, deviations can impair the control quality or even lead to substantial damage of the controlled system.
Hence, there is a strong need for a continuous validation of the system behavior during operation to enable quick reactions to any system changes.
However, the increasing data volumes in industrial systems pose large challenges for any such validation as the data needs to be transmitted and processed.

Addressing this pain point, we propose \oursystem{}, a \oursystemlongbf{} using \acf{INC}.
\oursystem{} validates the system behavior based on instantaneous and long-term process information which it checks for known dependencies using dedicated validation logic modules.
We demonstrate \oursystem{}’s efficacy by applying it to a lab-scale water treatment plant and show in our evaluation that \oursystem{} is capable of accurately detecting different failure states which can then be used to raise alerts or trigger a plant reconfiguration.
Overall, \oursystem{} shows that \ac{INC} can be sensibly deployed for supporting process control.

\section*{Acknowledgment}
Funded by the Deutsche Forschungsgemeinschaft (DFG, German Research Foundation) under Germany's Excellence Strategy – EXC-2023 Internet of Production – 390621612.

\bibliographystyle{IEEEtran}
\bibliography{literature}

\end{document}